%% file: bwt.tex
\documentclass[journal]{IEEEtran}

\usepackage{cite}
\usepackage{amsmath,amssymb,amsfonts,bm}
\usepackage{graphicx}
\usepackage{textcomp}
\usepackage{xcolor}
\usepackage{multicol}
\usepackage{booktabs}
\usepackage{multirow}
\usepackage{gensymb}
\usepackage{algorithm}
\usepackage{listings}

\usepackage{soul}
\usepackage{graphicx}
\usepackage{amsmath}
\usepackage{booktabs}
\usepackage{threeparttable}
\usepackage[hidelinks]{hyperref}

\input{macros}
\begin{document}

\title{Hardware Architecture for Inplace Compute of Burrows-Wheeler Transform in a Single Iteration}

\author{\IEEEauthorblockN{Kleber Stangherlin, Andrew Kennings} \\
        \IEEEauthorblockA{ECE Department, University of Waterloo, Waterloo, ON N2L 3G1, Canada}\\
        akennings@uwaterloo.ca
}

%\markboth{Submitted to IEEE for possible publication. Copyright may be transferred. This version may no longer be accessible.}{}

\maketitle

\begin{abstract}
The Burrows-Wheeler transform (BWT) is used by the bzip2 family of compressors. In this paper, we present a hardware architecture that implements an inplace algorithm to compute the BWT. Our design does not have explicit storage for the suffix array, or output array. The performance of our implementation is fixed, and does not depend on the input string content. We use a register based character buffer in a scanchain configuration, such that the BWT is computed from right to left, as characters are loaded. Loading new characters is done every six cycles, producing a new output character from the previously computed block at the same rate. Our FGPA implementation does not use block ram instances, and achieves throughput of 66, 35, 18, and 15~MB/s for block sizes of 128~B, 1~kB, 4~kB, and 8~kB. We also report results for an ASIC implementation in 65~nm CMOS that achieves 161~MB/s when using block size of 128~B.
\end{abstract}


\section{Introduction}

The Burrows-Wheeler transform (BWT) was introduced in~\cite{algoBwt1994}, and it is used by the \texttt{bzip2} family of compressors~\cite{resBzip2}. The BWT processes a block of text as a single unit, producing a new block of text that contains the same characters, but is easier to compress. The transform tends to group same characters together in the output, generating data that is easy to compress with locally-adaptive algorithms. First, the BWT algorithm performs $N$ rotations (cyclic-shifts) of the input string $S$ with length $N$. The rotated strings are then sorted lexicographically, and the transformed output $L$ is formed from the last character of each rotated string after sorting, as shown in Table~\ref{tab:intro}. Since the first and last characters of each rotated string are adjacent in the string $S$, consecutive characters in $L$ will also be adjacent to similar strings in $S$. Therefore, if the context of a character is a good predictor for the character, $L$ is likely to exhibit repeated characters, forming a string that is easier to compress.

Interest in hardware implementations of the BWT has risen in the recent years~\cite{relAss2019, relBitonicSorting2017, relLinearSorting2016}. Most techniques that implement BWT in hardware explore the fact that characters in $L$ are prefix to the first character of the corresponding rotated string. Therefore, finding the first column $F$, of the sorted suffix array (shifted strings) leads to the transformed output. If there are no repeated characters in $S$, the string $F$ is a sorted version of $S$. In the case of repeated characters in $S$, multiple sorting iterations are required where repeated characters are replaced by their suffix. While this approach reduces memory requirements, it makes performance dependent on the input data, since the occurrence of repeated characters will trigger additional sorting iterations.

In~\cite{algoBwtInplace2015}, authors introduced an algorithm to compute the BWT without explicit storage for the suffix array, or output array. The algorithm is shown in Listing~\ref{lst:inplace}. It relies on combinatorial properties of the BWT, and runs in $O(n^2)$ using $O(1)$ extra memory cells, apart from the memory required to store the $N$ characters of the input string. In this paper, we present a hardware architecture that implements such an algorithm in a single iteration, within a fixed number of clock cycles. The performance of our architecture is constant, and does not depend on the input string content. FPGA implementation results shown that our design achieves throughput of 66, 35, 18, and 15~MB/s for block sizes of 128~B, 1~kB, 4~kB, and 8~kB. The source code is publicly available in~\cite{resCodeBWT}.

This paper is organized as follows. In the next section, we discuss relevant works in the literature, highlighting their strengths and key differences from our work. Software implementations of the BWT are discussed next, with performance numbers on the \texttt{bzip2} implementation, and the inplace BWT algorithm in~\cite{algoBwtInplace2015}. A description of the hardware architecture for inplace compute of BWT is detailed next. The experimental results section reports area and performance numbers obtained from our architecture. This section also offers a comparison with prior works. Finally, this work ends with a conclusion, which also highlights possible points of improvements of the proposed hardware implementation.

\begin{table}[t]
    \centering
    \caption{Burrows-Wheeler transform.}
    \label{tab:intro}
    \begin{threeparttable}
        \input{tab_intro}
        \begin{tablenotes}[para,flushleft]
            Notes: Input to BWT is \texttt{banana\$}, and output is \texttt{annb\$aa}, where \texttt{\$} denotes end of text.
        \end{tablenotes}
    \end{threeparttable}
\end{table}

\begin{table*}[t]
    \centering
    \caption{BZIP2 profiling with various corpus and block sizes.}
    \label{tab:bzip2}
    \begin{threeparttable}
        \input{tab_bzip2}

        \begin{tablenotes}[para,flushleft]
            Notes: Average values from corpus \texttt{E.coli.txt}, \texttt{bible.txt}, \texttt{book1.txt}, \texttt{book2.txt}, and \texttt{world192.txt}. Allocated memory is 8x the block size, and it is used for both BWT and compression.
        \end{tablenotes}
    \end{threeparttable}
\end{table*}

\section{Related Works}

The first attempt to implement BWT in hardware was done in~\cite{relWeavesorter2001}, where the input string is right shifted and compared at each step. The sorted string is obtained by left shifting the stored data. This method is known as Weavesorter, and it requires software replacement of repeated values in the string between sorting iterations. A parallel sorting approach that requires $N/2$ steps was proposed in~\cite{relParallelSorting2005}. It is faster than Weavesorter, but still requires the replacement of repeated characters, followed by another sorting iteration which only sorts data that has changed.

In~\cite{relLcp2013}, authors introduce an architecture to compute BWT in only one iteration. Their implementation saves the context of each character and performs the sorting operations on them. The size of saved contexts is given by the longest common prefix in the string. Therefore, synthesizing the hardware requires upfront knowledge of the input string content, which reduces the applicability of the proposed architecture.

A linear sorter architecture is proposed in~\cite{relLinearSorting2016}, where the input is sorted as it is loaded. Replacement of repeated characters and additional sorting iterations are still required, but only the required characters are loaded and sorted again. In~\cite{relBitonicSorting2017}, authors introduce an architecture that uses a bitonic sorting network to compute the BTW with block size up to 4~kB. They increase throughput by computing multiple blocks in parallel, but extra sorting operations that depend on input string content are still required.

An implementation capable of computing BWT with block size of 500~kB was proposed in~\cite{relAss2019}. It uses a large number of BRAMs to reduce the cache misses of an antisequential suffix sorting algorithm introduced in~\cite{relAssDetails2005}. Authors report an average speedup of 2x, compared to the CPU implementation in the \texttt{bzip2} software package.

\begin{table}[t]
    \centering
    \caption{Original BWT paper results~\cite{algoBwt1994}.}
    \label{tab:bwt}
    \begin{threeparttable}
        \input{tab_bwt}
        \begin{tablenotes}[para,flushleft]
            Notes: \texttt{book1.txt} from Calgary compression corpus, and the Hector corpus~\cite{resHector1992}. Compression uses move-to-front algorithm and a modified Huffman coding~\cite{algoBwt1994}.
        \end{tablenotes}
    \end{threeparttable}
\end{table}

\section{Burrows-Wheeler Transform}

The Burrows-Wheeler transform (BWT) reorganizes the characters in a text block such that the transformed text tend to have same characters grouped together, which enables more efficient data compression~\cite{algoBwt1994}.

\subsection{Software Implementation}

\input{lst_bwt}

\begin{table*}[t]
    \centering
    \caption{Performance results for software implementation of inplace algorithm to compute BWT~\cite{algoBwtInplace2015}.}
    \label{tab:inplace}
    \begin{threeparttable}
        \input{tab_inplace}
        \begin{tablenotes}[para,flushleft]
            Notes: corpus \texttt{bible.txt}, with 4047392 bytes.
        \end{tablenotes}
    \end{threeparttable}
\end{table*}

A naive software implementation of the BWT algorithm takes $O(N^2\log N)$ time and requires $O(N^2)$ space. We profiled the reference implementation of \texttt{bzip2}, available in~\cite{resBzip2}, with various corpus and different block sizes. Results are listed in Table~\ref{tab:bzip2}, including values for compression ratio (bits per character), percentage of time spent in BWT over the total execution time, BWT routine throughput, and allocated memory for the entire operation. Compression results for smaller text block sizes are listed in Table~\ref{tab:bwt}, which was extracted from the original BWT publication~\cite{algoBwt1994}.

The variation in compression rate reported in Table~\ref{tab:bzip2} in the 100~kB to 900~kB block size range is 14\%. Alternatively, Table~\ref{tab:bwt} shows less effective compression rates, most likely due to different compression algorithms used. The average improvement in compression rate for the 1~kB to 4~kB range is 11\%. Another 11\% improvement in compression rate is obtained when block size is increased from 4~kB to 16~kB.

The importance of BWT performance is also noticeable from results in Table~\ref{tab:bzip2}. The BWT routine takes more than 60\% of \texttt{bzip2} execution time for all text block sizes. The measured throughput is nearly constant at 12~MB/s, while the allocated memory for the whole operations was 8x the block size used.

\subsection{Inplace Algorithm with Single Iteration}

The algorithm in Listing~\ref{lst:inplace} was introduced in~\cite{algoBwtInplace2015} to compute the BWT without explicit storage for the suffix array, or the output array. It computes the BWT from right to left, in $O(N^2)$ steps, using $O(1)$ extra memory cells, apart from the cells storing the input text block. Table~\ref{tab:bwt} shows performance information of our CPU implementation. Throughput quickly degrades for large block sizes, following the expected quadratic time complexity. In the next sections, we show that careful implementation of the algorithm in hardware leads to a substantial increase in throughput.

\section{Hardware Implementation}

\begin{figure}[t]
    \centering
    \includegraphics[scale=1]{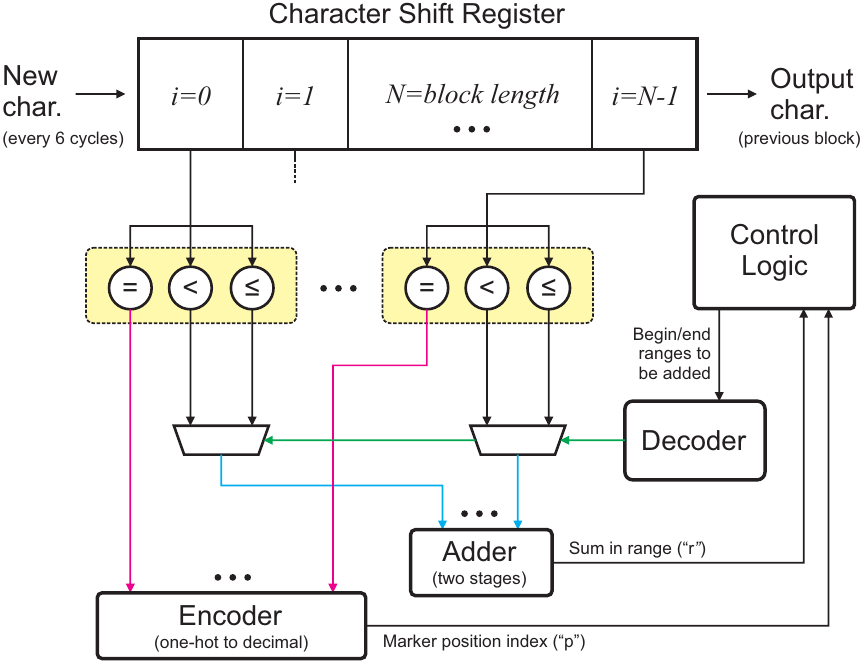}
    \caption{Hardware architecture to compute the Burrows-Wheeler transform with $O(1)$ extra memory.}
    \label{fig:bwarch}
\end{figure}

Our hardware architecture for inplace compute of BWT is shown in Fig.~\ref{fig:bwarch}. The transformed output is computed in a fixed number of steps, independent of the input string content. Every 6 cycles a new character is loaded, and another character from the previous block is made available at the output. The character buffer is implemented with registers in a scanchain configuration, such that it fits a complete text block. The algorithm proceeds from right to left, performing the BWT of a gradually increasing text block size, until the transform of the entire block is computed. The comparison signals necessary for sorting characters are generated in parallel for all characters in the buffer. An encoder finds the end marker position in the character buffer, which is then used to define the ranges to compute \texttt{r} (steps 1 and 2 in the algorithm).

\begin{figure}[t]
    \centering
    \includegraphics[scale=1]{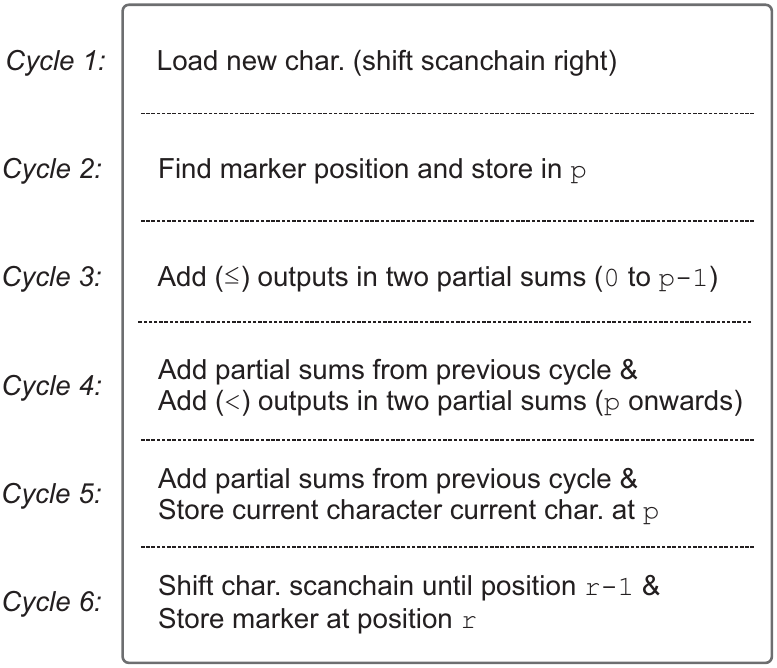}
    \caption{Operations performed in each of the 6 cycles used by our hardware implementation.}
    \label{fig:cycles}
\end{figure}

Every iteration of the outer loop in Listing~\ref{lst:inplace} takes 6 cycles, as shown in Fig.~\ref{fig:cycles}. Cycle 1 simply loads the new character and shifts the scanchain; as a side effect, a new output character from the previously computed block is produced. Cycle 2 uses the equality comparators to find the end marker position, which is used in cycle 3 to perform a population count over the ($\leq$) comparison results. Similarly to cycle 3, cycle 4 also performs a population count, but on the ($<$) comparison results. Cycles 5 and 6 perform repositioning of the newly added character and the end marker in the buffer.

Timing analysis revealed that the population count in cycles 3 and 4 has substantial impact on the maximum frequency. Therefore, we split the addition over two cycles by computing the result in two partial sums that are added together in the coming cycle. This change increased the maximum frequency up to 48\%, depending on block size.

\section{Experimental Results}

\begin{table*}[t]
    \centering
    \caption{Experimental results and comparison with other Burrows-Wheeler hardware architectures.}
    \label{tab:cmp}
    \begin{threeparttable}
        \input{tab_cmp}
        \begin{tablenotes}[para,flushleft]
            Notes: (*) Complete part number XCVU9P-L2FSGD2104E.  (\textdagger) Complete part number EP1S10B672C6. (\textdaggerdbl) Imposes restrictions to input string content. (\textdagger\textdagger) Multiple blocks are computed in parallel.
        \end{tablenotes}
    \end{threeparttable}
\end{table*}

Our hardware implementation is publicly available in~\cite{resCodeBWT}. This section presents area and performance results for FPGA and 65~nm CMOS targets. 

\subsection{FPGA Implementation}

Our FPGA implementation was synthesized using Synopsys Synplify tool. Our target device was an FPGA with part number XCVU9P-L2FSGD2104E. Placement and routing was done using standard Xilinx tools. Results for various block sizes are reported in Table~\ref{tab:cmp}, including a comparison with other recent works.

As shown in Table~\ref{tab:cmp}, our architecture uses no block rams (BRAMs). Our storage requirements are simply the registers necessary to store the input text block. This is only matched by the work in~\cite{relLinearSorting2016}, however, our design not only reaches almost twice the throughput, but does so in a single iteration. Moreover, compared to~\cite{relLinearSorting2016}, we support much larger text block sizes. The best results in terms of throughput were reported by~\cite{relBitonicSorting2017}, however, it is important to notice that authors used several instances to compute multiple blocks simultaneously, which is equally applicable to our design.

\subsection{ASIC Implementation}

\begin{figure}[t]
    \centering
    \includegraphics[scale=1.1]{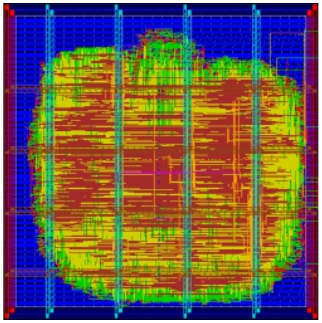}
    \caption{ASIC implementation of inplace BWT accelerator in 65~nm CMOS.}
    \label{fig:chip}
\end{figure}

Our ASIC implementation uses a 128~B block size and targets a 65~nm CMOS process. The synthesis and physical implementation were done using the Genus and Innovus. The accelerator layout is shown in Fig.~\ref{fig:chip}. It uses a square layout with 380~$\mu m$ side length, while the actual design area uses 19260 instances and 92680~$\mu m^2$. The maximum frequency 843~MHz, which is capable of 161~MB/s.

\section{Conclusion}

We presented a hardware implementation for inplace compute of the BWT. Our design does not have explicit storage for the suffix array, or output array. We do not use any block ram memory. The performance is fixed, and does not depend on the input string content. Our area and performance results are very competitive for small block sizes like 128~B, and 1~kB. Future work could potentially increase throughput by computing multiple blocks in parallel. Moreover, improvements to the performance for large block sizes are likely possible by redesigning the comparators to generate results over multiple cycles.

\bibliographystyle{plain}
\bibliography{bwt}

\end{document}

%% file: tab_intro.tex
\begin{tabular}{@{}lll@{}}
\toprule
\textbf{Cyclic shifts} & \textbf{Sorted shifts} & \textbf{Output} \\ \midrule
\texttt{banana\$}               & \texttt{\$banan\textbf{a}}               & \texttt{a}               \\
\texttt{\$banana}               & \texttt{a\$bana\textbf{n}}               & \texttt{n}               \\
\texttt{a\$banan}               & \texttt{ana\$ba\textbf{n}}               & \texttt{n}               \\
\texttt{na\$bana}               & \texttt{anana\$\textbf{b}}               & \texttt{b}               \\
\texttt{ana\$ban}               & \texttt{banana\textbf{\$}}               & \texttt{\$}              \\
\texttt{nana\$ba}               & \texttt{na\$ban\textbf{a}}               & \texttt{a}               \\
\texttt{anana\$b}               & \texttt{nana\$b\textbf{a}}               & \texttt{a}               \\ \bottomrule
\end{tabular}

%% file: tab_bzip2.tex
\begin{tabular}{@{}llllllllll@{}}
\toprule
\textbf{Block size}    & 100~kB    & 200~kB    & 300~kB    & 400~kB    & 500~kB    & 600~kB    & 700~kB    & 800~kB    & 900~kB    \\ \midrule
\textbf{Bits per char} & 2.31~bits & 2.18~bits & 2.11~bits & 2.07~bits & 2.05~bits & 2.02~bits & 2.01~bits & 1.98~bits & 1.98~bits \\
\textbf{BWT Percent.}  & 60.1\%    & 64.1\%    & 62.2\%    & 70.0\%    & 66.2\%    & 67.4\%    & 73.4\%    & 70.0\%    & 69.7\%    \\
\textbf{Throughput}    & 12~MB/s   & 12~MB/s   & 14~MB/s   & 12~MB/s   & 11~MB/s   & 12~MB/s   & 11~MB/s   & 12~MB/s   & 12~MB/s   \\
\textbf{Alloc. memory}   & 0.78~MB   & 1.56~MB   & 2.34~MB   & 3.12~MB   & 3.90~MB   & 4.68~MB   & 5.47~MB   & 6.25~MB   & 7.03~MB   \\ \bottomrule
\end{tabular}

%% file: tab_bwt.tex
\begin{tabular}{@{}lll@{}}
\toprule
\textbf{}           & \multicolumn{2}{c}{\textbf{Bits per character}} \\ \cmidrule(l){2-3} 
\textbf{Block size} & \texttt{book1.txt}   & \texttt{Hector}  \\ \midrule
1~kB                & 4.34             & 4.35                    \\
4~kB                & 3.86             & 3.83                    \\
16~kB               & 3.43             & 3.39                    \\
64~kB               & 3.00             & 2.98                    \\
256~kB              & 2.68             & 2.65                    \\
750~kB              & 2.49             & --                      \\
1~MB                & --               & 2.43                    \\
4~MB                & --               & 2.26                    \\
16~MB               & --               & 2.13                    \\
64~MB               & --               & 2.04                    \\
103~MB              & --               & 2.01                    \\ \bottomrule
\end{tabular}

%% file: lst_bwt.tex
\begin{lstlisting}[float=t,caption={Inplace algorithm to compute BWT, introduced in~\cite{algoBwtInplace2015}.}, captionpos=b,label={lst:inplace}]
void inplaceBWT( unsigned char S[ ], int n ){
  int i, p, r, s;
  unsigned char c;
  for ( s = n-3; s >= 0; s-- ){
    c = S[ s ];
    /* steps 1 and 2 */
    r = s;
    for ( i = s+1; S[ i ] != END_MARKER; i++ )
      if ( S[ i ] <= c ) r++;
    p = i;
    while ( i < n )
      if ( S[ i++ ] < c ) r++;
    /* step 3 */
    S[ p ] = c;
    /* step 4 */
    for ( i = s; i < r; i++ )
      S[ i ] = S[ i+1 ];
    S[ r ] = END_MARKER;
  }
}
\end{lstlisting}

%% file: tab_inplace.tex
\begin{tabular}{@{}llllllllll@{}}
\toprule
\textbf{Block size}       & 1~kB     & 2~kB     & 4~kB      & 8~kB      & 16~kB     & 32~kB     & 64~kB     & 128~kB    & 256~kB    \\ \midrule
\textbf{Number of blocks} & 3953     & 1977     & 989       & 495       & 248       & 124       & 62        & 31        & 16        \\
\textbf{Compression time} & 1.4~s    & 2.5~s    & 4.9~s     & 9.6~s     & 19~s      & 38~s      & 76~s      & 152~s     & 300~s     \\
\textbf{Throughput}       & 2.8~MB/s & 1.5~MB/s & 0.78~MB/s & 0.40~MB/s & 0.20~MB/s & 0.10~MB/s & 0.05~MB/s & 0.02~MB/s & 0.01~MB/s \\ \bottomrule
\end{tabular}

%% file: tab_cmp.tex
\begin{tabular}{@{}l|llll|llll@{}}
\toprule

& \multicolumn{4}{c|}{\multirow{2}{*}{\textbf{This work}}}
& \textbf{FCCM'19}
& \textbf{ICFPT'17}
& \textbf{IPDPSW'16}
& \textbf{ReConFig'13} \\

& \multicolumn{4}{c|}{}
& \multicolumn{1}{c}{\cite{relAss2019}}
& \multicolumn{1}{c}{\cite{relBitonicSorting2017}}
& \multicolumn{1}{c}{\cite{relLinearSorting2016}}
& \multicolumn{1}{c}{\cite{relLcp2013}} \\ \midrule

\textbf{Device}         & XCVU9P*        & XCVU9P*        & XCVU9P*        & XCVU9P*      & XCVU9P*    & XC7VX690T   & XC7K70T       & EP1S10\textsuperscript{\textdagger}     \\
\textbf{Technology}     & 20~nm          & 20~nm          & 20~nm          & 20~nm        & 20~nm      & 28~nm       & 28~nm         & 0.13~um      \\
\textbf{Block size}     & 128            & 1~kB           & 4~kB           & 8~kB         & 300~kB     & 4~kB        & 128~bits      & 1~kB         \\
\textbf{BRAMs}          & 0\%            & 0\%            & 0\%            & 0\%          & 21.4\%     & --          & 0\%           & 64~kB        \\
\textbf{Registers}      & 1.1~k (0.05\%) & 8.7~k (0.36\%) & 34~k (1.4\%)   & 67~k (2.8\%) & --         & --          & --            & 10~k         \\
\textbf{LUTs (logic)}   & 3.5~k (0.3\%)  & 29.7~k (2.5\%) & 121~k (10.3\%) & 281~k (24\%) & 1\%        & --          & 5.4~k (11\%)  & --           \\
\textbf{Inplace}        & Yes            & Yes            & Yes            & Yes          & No         & No          & Yes           & No           \\
\textbf{Max Freq.}      & 345~MHz        & 186~MHz        & 95~MHz         & 69~MHz       & 187~MHz    & 155~MHz     & 152~MHz       & 130~MHz      \\
\textbf{Throughput}     & 66~MB/s        & 35~MB/s        & 18~MB/s        & 15~MB/s      & 13.4~MB/s  & 179~MB/s    & 37.1~MB/s     & 62~MB/s      \\
\textbf{Has Parallel.}  & No             & No             & No             & No           & No         & Yes (block)\textsuperscript{\textdagger\textdagger} & No            & No           \\
\textbf{\# of Cycles}   & 768            & 6144           & 24576          & 49152        & --         & 4049        & 500           & 2048         \\
\textbf{Iterations}     & Single         & Single         & Single         & Single       & Input Dep. & Input Dep.  & Input Dep.    & Single\textsuperscript{\textdaggerdbl}            \\
\textbf{Open-source}    & Yes            & Yes            & Yes            & Yes          & No         & No          & No            & No            \\ \bottomrule
\end{tabular}